\documentstyle[12pt]{article}

\begin{document}
\thispagestyle{empty}
\begin{center}
\LARGE \tt \bf{Criteria for de Sitter inflation in Einstein-Cartan cosmology and COBE data}
\end{center}
\vspace{1cm}
\begin{center} {\large L.C. Garcia de Andrade\footnote{Departamento de
Matematica Aplicada-IMECC=Universidade Estadual de Campinas,Brasil.
e-mail.: garcia@ime.unicamp.br.Permanent address:Departamento de Fisica Teorica-IF-UERJ-CEP:20550-013,Rio de Janeiro,RJ}}
\end{center}
\vspace{1.0cm}
\begin{abstract}
Criteria for the existence of de Sitter inflation with dilaton fields in four-dimensional space-times with torsion is discussed.The relation between matter density perturbation and the spin-density perturbation is stablished based on this criteria.From COBE data it is shown that there is a linear relationship between the spin-torsion density and temperature of the Universe for the case where matter density dominates the kinetic part of dilaton fields.
\end{abstract}      
\vspace{1.0cm}       
\begin{center}
\Large{PACS number(s) : 0420,0450,1127}
\end{center}
\newpage
\pagestyle{myheadings}
\markright{\underline{Criteria for de Sitter inflation in Einstein-Cartan cosmology and COBE data}}
\paragraph*{}
Recently much work has been done in Inflationary theories in Einstein-Cartan cosmology \cite{1,2,3,4,5} without avery clear criteria of how to define inflation with dilaton fields or inflatons.In this letter we hope to remedy this situation by stablishing this criteria following the same pattern that was used in the context of General Relativity (GR) \cite{6}.Application of this criteria to an example where we investigate the relation between the spin-torsion and matter densities and the temperature anisotropies is given, where the COBE data is used to obtain an expression between the spin-torsion density and temperature of the Universe.This letter complements previous work developed by Palle \cite{4} in the sense that we deal with inflation generated by a scalar field which is not taken into account when Palle investigates a model of inflation with rotation and expansion and computes the density perturbation during this phase.Korotky \cite{5} has also applied this model to show that is possible to solve some problems that appear in the Nucleosynthesis in Riemann-Cartan spacetime.To start with let us consider the Einstein-Cartan equations as given in Gasperini \cite{1}
\begin{equation}
H^{2}=\frac{8{\pi}G}{3}({\rho}_{eff}-2{\pi}G{\sigma}^{2}) 
\label{1}
\end{equation}
and
\begin{equation}
H^{2}=-\frac{4{\pi}G}{3}({\rho}_{eff}+3{p}_{eff}-8{\pi}G{\sigma}^{2}) 
\label{2}
\end{equation}
Where $\frac{\dot{a}}{a}=H$.Application of the de Sitter metric
\begin{equation} 
ds^{2}=dt^{2} - e^{2Ht}(dz^{2}+dx^{2}+dy^{2})
\label{3}
\end{equation}
to the EC equations yields
\begin{equation}
{\rho}_{eff}=-p_{eff}+4{\pi}G{\sigma}^{2}
\label{4}
\end{equation}
Where
\begin{equation}
{\rho}_{eff}={\dot{\phi}}^{2}+V({\phi})+{\rho}
\label{5}
\end{equation}
and 
\begin{equation}
{p}_{eff}={\dot{\phi}}^{2}-V({\phi})+{\rho}
\label{6}
\end{equation}
Substitution of expressions (\ref{5}) and (\ref{6}) into expression (\ref{4}) yields after some algebra
\begin{equation}
2{\dot{\phi}}^{2}+{\rho}=-p+4{\pi}G{\sigma}^{2}
\label{7}
\end{equation}
One must note that this criteria reduces to the GR criteria when torsion and the dilaton field ${\phi}$ vanish.Expression (\ref{7}) gives us some more freedom on the criteria to choose to consider inflation in the context of Einstein-Cartan gravity.The first could be to maintain the criteria of GR inflation and to use ${\rho}=-p$ into expression (\ref{7}) yields
\begin{equation}
{\dot{\phi}}^{2}=2{\pi}G{\sigma}^{2}
\label{8}
\end{equation}
In the particular case where torsion is constant integration of expression (\ref{8}) yields a linear relation between dilaton and time where torsion appears in the angular coefficient of the straight lines as
\begin{equation}
{\phi}(t)={\phi}(0)+2{\pi}G{\sigma}_{0}t
\label{9}
\end{equation}
where the index $0$ indicates that the quantity is constant.By making use of the relation $a(t)=\frac{1}{T}$ between the cosmic scale factor $a$ and the temperature $T$,one obtains 
\begin{equation}
{\delta}a=-\frac{{\delta}T}{{T}^{2}}
\label{10}
\end{equation}
Now from the EC field equations above one obtains 
\begin{equation}
\frac{{\delta}{\rho}}{\rho}|_{eff}=-2k\frac{T{\delta}T}{H^{2}-kT^{2}}
\label{11}
\end{equation}
By making use of the approximation $kT^{2}>>>H^{2}$ one obtains the relationship between the effective matter density fluctuation and the temperature fluctuation
\begin{equation} 
\frac{{\delta}{\rho}}{\rho}|_{eff}=2\frac{{\delta}T}{T}
\label{12}
\end{equation}
Yet from the field equations one now has 
\begin{equation}
\frac{{\delta}{\rho}}{\rho}|_{eff}=\frac{2}{T}\frac{{\delta}{\sigma}}{\sigma}-3\frac{k}{16{\pi}G^{2}{\sigma}^{2}}2\frac{{\delta}T}{T}
\label{13}
\end{equation}
To simplify matters we shall consider the flat section $(k=0)$ and the last term drops out.From the expression (\ref{5}) one obtains 
\begin{equation}
\frac{{\delta}{\rho}}{\rho}=\frac{1}{T}\frac{{\delta}{\sigma}}{\sigma}
\label{14}
\end{equation}
and from the COBE data $\frac{{\delta}T}{T}=10^{-5}$ we obtain a relation between the spin-torsion density fluctuation and the temperature of the Universe as 
\begin{equation}
\frac{{\delta}{\sigma}}{\sigma}=\frac{1}{2}.10^{-5}T
\label{15}
\end{equation}
Just to give an estimate to the spin-torsion density during the de Sitter inflationary phase let us consideer the temperature $T_{0}=0.5 K$ for the temperature of the CBR radiation \cite{7}which yields
\begin{equation}
\frac{{\delta}{\sigma}}{\sigma}=2.5 10^{-6}
\label{16}
\end{equation}
In principle an experiment could be proposed to measured such relative high value predicted for the spin-torsion density fluctuation based on the Einstein-Cartan cosmology.
\section*{Acknowledgments}
\paragraph*{}
Thanks are due to Prof.I.L.Shapiro and Prof.Rudnei Ramos, for their constant advice on the subject of this paper.I am very much indebt to FAPESP (fundacao de Amparo a Pesquisa do Estado de Sao Paulo) and CNPq. (Brazilian Government Agency) for financial support.


\begin{thebibliography}{7}
\bibitem{1}M.Gasperini,Repulsive gravity in the very early universe,gr-qc/9805060.
\bibitem{2}L.C.Garcia de Andrade,Phys.Lett.B468,(1999)28.
\bibitem{3}A.Maroto and I.Shapiro,Phys.Lett.B 414,(1997)34.
\bibitem{4}D.Palle,On Primordial Cosmological Density Fluctuations in the Einstein-Cartan Gravity
from COBE data,gr-qc LOs Alamos archives (1999).
\bibitem{5}A.Korokty,in Modern Problems of theoretical Physics,Fetschrift for D.Ivanenko,(1990),World Scientific. 
\bibitem{6}R.Brandeenberger,in Gravitation and Cosmology Brazilian School,(1995),Ed. Mario Novello.
\bibitem{7}J.Peebles,Principles of Physical Cosmology,(1993),Princeton University Press.
\end{thebibliography}
\end{document}